# Deconvolution of phonon scattering by ferroelectric domain walls and point defects in a PbTiO$_3$ thin film deposited in a composition-spread geometry

David Bugallo[1], Eric Langenberg[1,2,3,*], Elias Ferreiro-Vila[1], Eva H. Smith[2], Christina Stefani[4], Xavier Batlle[3], Gustau Catalan[4], Neus Domingo[4], Darrell G. Schlom[2,5,6], and Francisco Rivadulla[1,*]

[1]CiQUS-Universidade de Santiago de Compostela, Santiago de Compostela 15782, Spain
[2]Department of Materials Science and Engineering, Cornell University, Ithaca, New York 14853, USA
[3]Department de Física de la Matèria Condensada and Institut de Nanociència i Nanotecnologia (IN2UB), Universitat de Barcelona, Barcelona 08028, Spain
[4]Catalan Institute of Nanoscience and Nanotechnology (ICN2), CSIC, Barcelona Institute of Science and Technology, Campus Universitat Autònoma de Barcelona, Bellaterra, 08193 Barcelona, Spain
[5]Kavli Institute at Cornell for Nanoscale Science, Ithaca, New York 14853, USA
[6]Leibniz-Institut für Kristallzüchtung, Max-Born-Str. 2, 12489 Berlin, Germany

**We present a detailed analysis of the temperature dependence of the thermal conductivity of a ferroelectric PbTiO$_3$ thin film deposited in a composition-spread geometry enabling a continuous range of compositions from ~25% titanium-deficient to ~20% titanium-rich to be studied. By fitting the experimental results to the Debye model we deconvolve and quantify the two main phonon scattering sources in the system: ferroelectric domain walls (DWs) and point defects. Our results prove that ferroelectric DWs are the main agent limiting the thermal conductivity in this system, not only in the stoichiometric region of the thin film ([Pb]/[Ti]≈1), but also when the concentration of cation point defects is significant (up to ≈15%). Hence, DWs in ferroelectric materials are a source of phonon scattering at least as effective as point defects. Our results demonstrate the viability and effectiveness of using reconfigurable DWs to control the thermal conductivity in solid-state devices.**

Keywords: composition-spread combinatorial study, ferroelectrics, thermal conductivity, phonon scattering mechanisms, domain walls

# INTRODUCTION

The possibility of dynamically controlling the thermal conductivity in solid-state devices using ferroelectric materials[1–7] is an exciting emerging research direction.[8,9] At the heart of this process is the nature of the ferroelectric domain walls (DWs): they may be very efficient phonon scattering centers,[2–4,10–12] and at the same time, they can be moved, created, and erased with an electric field,[13–15] providing a unique opportunity to electrically modulate phonon transport. This would enable the development of thermal transistors and thermal switches, key elements of effective thermal circuitry, basic phononic logic operations, and novel devices in thermal energy management and harvesting. In this regard, substantial progress has been produced in polycrystalline $PbZr_{0.3}Ti_{0.7}O_3$ films, in which the thermal conductivity was varied by 11% at room temperature by switching on and off the applied voltage.[1] Subsequent theoretical reports predicted that a much larger modulation could be feasible in single-crystal ferroelectric materials.[3,4,6] Recently, the thermal conductivity in epitaxially-grown $PbTiO_3$ films was shown to vary up to 60% at room temperature depending on the density of domain walls induced by strain engineering.[7]

In addition to the ferroelectric DWs, extrinsic interfaces (i.e., grain boundaries,[16,17] interface boundaries in multilayers[18,19]) and point defects (vacancies, atomic substitutions[20–23]) may also contribute to the scattering of phonons and affect the thermal conductivity. Importantly, oxygen vacancies, which are very common point defects in oxide materials, can be quite mobile under the presence of an electric field.[24–26] Specifically, these vacancies can be concentrated or depleted where the electric field is applied,[26] which may result in a substantial change in the thermal conductivity because of a modulation in the density of point defects. Thus, the electric-field-control of the phonon transport in ferroelectric materials might arise from the ferroelectric switching, from mobile point defects, or, more likely, from a combination of both. Yet, while the electrical modification of the domain wall patterns can be stable[27]—enabling stable "phononic" states—the fact that the ion vacancies gradually diffuse back to the original configuration once the electric field is removed[26] would create transitory "phononic" states. It is thus very important to disentangle both contributions to the phonon scattering and thermal conductivity in ferroelectric materials.

For this purpose, we have grown an ~80 nm thick ferroelectric $PbTiO_3$ thin film using reactive molecular-beam epitaxy (MBE) onto a 1"-diameter (001)-oriented $SrTiO_3$ single-crystal substrate in a composition-spread geometry enabling a continuous range of compositions from ~25% titanium-deficient to ~20% titanium-rich to be studied. The intentional gradient in the [Pb]/[Ti] relative composition along one of the diameters of the wafer (see the Suppl. Info.) arises due to the substrate not being rotated during film deposition. After growth, the sample was cut into seven pieces along this diameter. Each of the resulting $PbTiO_3$ pieces has a different [Pb]/[Ti] compositional ratio: three increasingly lead rich, three increasingly titanium rich, and one stoichiometric. Analyzing the thermal conductivity of these pieces over a wide temperature range allowed us to investigate the interplay between point defects, ferroelectric domain walls, and phonon transport, as discussed next.

**RESULTS AND DISCUSSION**

The progressive departure from the stoichiometric [Pb]/[Ti] ratio ([Ti]/[Ti]$_{Stoich.}$ = 1, where the titanium concentration measured by energy dispersive X-ray spectroscopy (EDS) is normalized by that of the stoichiometric film as described in the Suppl. Info.) produces a gradual increase in the $2\theta$ values of the X-ray diffraction (XRD) 002 Bragg peak of the $PbTiO_3$ film (see

Figure S2a in the Suppl. Info.). The subsequent contraction of the out-of-plane lattice parameter is consistent with a partial relaxation of the compressive epitaxial strain through vacancy creation, as observed in other perovskites oxides.[28,29] Given the large impact that point defects have on phonon scattering,[20–23] one would expect a monotonic suppression of the thermal conductivity, $\kappa$, of $PbTiO_3$ as the composition departs from stoichiometric.

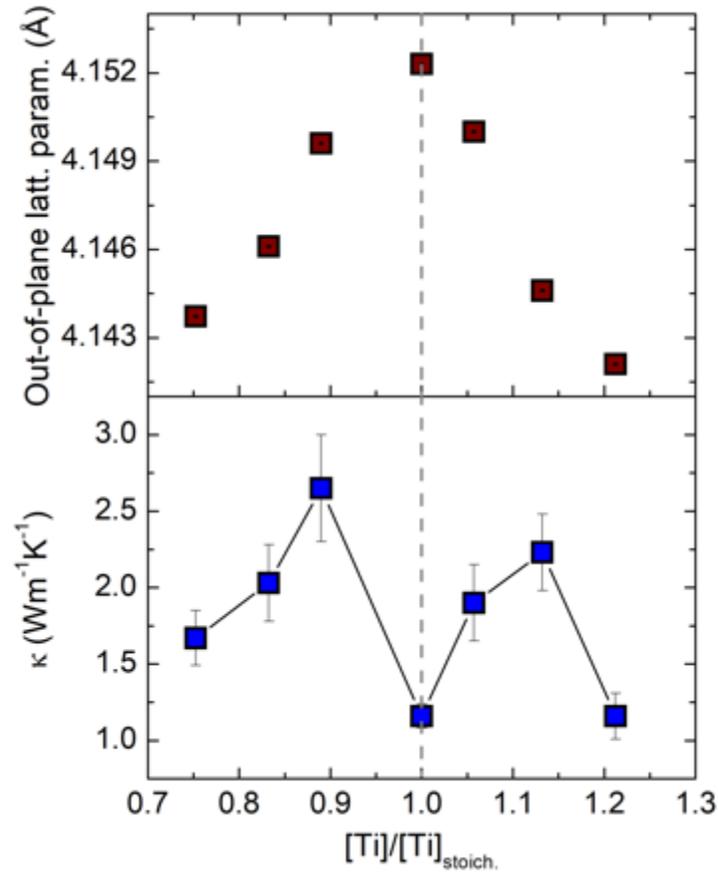

**Figure 1.** Out-of-plane lattice parameter (top panel) and cross-plane thermal conductivity (bottom panel, measured at 250 K) of a $PbTiO_3$ film deposited in a composition spread geometry as a function of $[Ti]/[Ti]_{stoich.}$ (i.e., the titanium concentration normalized by that of the stoichiometric film). The temperature range of our 3$\omega$-method device is limited to $\approx$290 K. To avoid problems with the calibration, we plot the value of $\kappa$ at 250 K, although a similar result is obtained between 100 K and 275 K, where $\kappa$ does not change much with temperature (see Figure 3 below).

Yet, surprisingly enough, a quite different behavior is found (Figure 1, bottom panel). The most stoichiometric film displays about the lowest $\kappa$, which increases as the $[Ti]/[Ti]_{Stoich.}$ departs from 1, either to the lead-rich ($[Ti]/[Ti]_{Stoich} <1$) or titanium-rich region ($[Ti]/[Ti]_{Stoich} >1$), until reaching a maximum. Further departure from the cationic stoichiometry results in reduction of $\kappa$. So, it is evident that the monotonic increase of the concentration of point defects alone cannot explain the thermal conductivity in the ferroelectric film deposited in a composition-spread geometry.

As aforementioned, DWs in ferroelectric materials may also have a dramatic impact on the thermal conductivity as well.[7] The evolution of the ferroelectric domain patterns with composition was studied on the same film by vertical piezoresponse force microscopy (PFM)—see the experimental details in the Suppl. Info. The results are summarized in Figure 2a,b. The most stoichiometric sample displays a mixture of *c*-domains with up and down polarizations, yet with one type of domain clearly favored (note the abundance of one type of polarization in the phase image in Figure 2a). A preferential orientation of *c*-domains in stoichiometric $PbTiO_3$ films grown on $SrTiO_3$ was previously observed,[7] related to a built-in electric field arising from the surface of the substrate.[14] In this case, however, the *c*-domains form a striking pattern with a worm-like shape. This unprecedented morphology may come from some cation inhomogeneities (excess or deficit of either lead or titanium) occurring at the off-stoichiometric extremes which may diffuse towards the center of the sample during growth, adopting the squiggly form observed. This hypothesis also implies the existence of a threshold in the [Pb]/[Ti] ratio above/below which the polarization of the *c*-domains will show one direction or the opposite. Indeed, the up or down orientation of the polarization of the *c*-domains in $PbTiO_3$ films has recently been found to be dependent on the growth temperature, regardless of the electrostatic boundary condition: up for low temperatures, down for high temperatures, and a mixture of up and down for intermediate temperatures.[30] Assessing the lead content in the films, these authors proved that this behavior is related to the excess (deficit) of lead at low (high) growth temperatures due to the volatility of lead.[30] Relevant here is the fact that, due to this phenomenon, the density of the worm-like domains gradually decreases on departing from stoichiometric $PbTiO_3$ (Figure 2b) as the samples become monodomain up (excess of lead) or down (deficit of lead). This evolution is also observed in the reciprocal space maps around 002 XRD peaks (Figure 2c): the $PbTiO_3$ diffraction spots become sharper with increasing departure from stoichiometry (see also Suppl. Info.), indicating the gradual decrease in the DW density, as reported previously.[7] This trend is confirmed by the narrowing of the FWHM of rocking curves (Figure S2b Suppl. Info.) around the 001 XRD reflections of the $PbTiO_3$ film when the density of DWs (point defects) decreases (increases). These results corroborate the tendency revealed locally by the PFM results (Figure 2a,b) at the macroscopic scale: the DW density decreases on increasing the concentration of cationic defects.

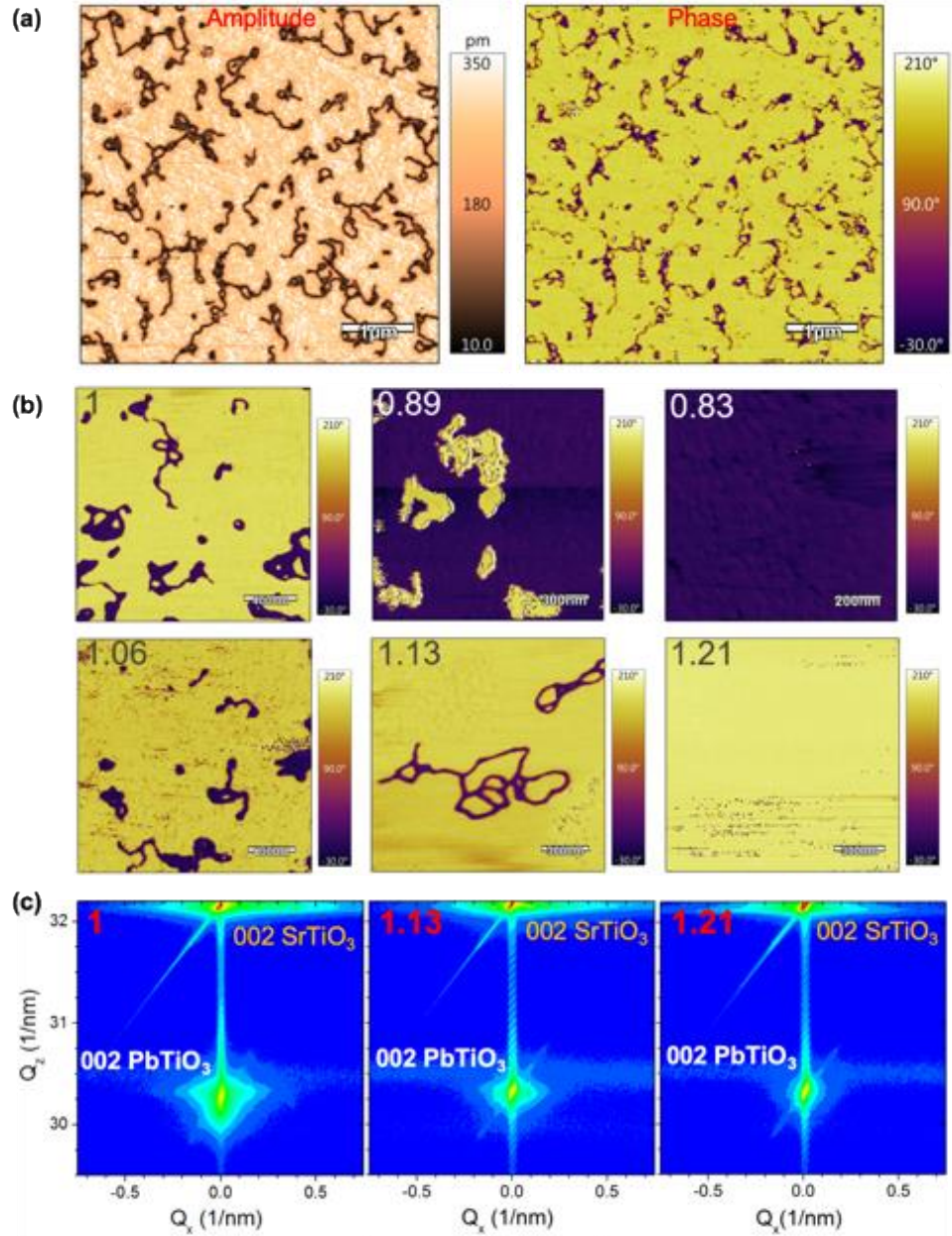

**Figure 2. (a)** Amplitude (left panel) and phase (right panel) vertical PFM images of the stoichiometric piece of the PbTiO$_3$ film. **(b)** Evolution of the vertical PFM phase with the titanium content in the PbTiO$_3$ film deposited in a composition-spread geometry. **(c)** XRD reciprocal space maps around the 002 reflection of the PbTiO$_3$ film as a function of the titanium content. The number at the top left corner of the images in (b) and (c) indicates the normalized [Ti]/[Ti]$_{Stoich.}$ values.

On the other hand, stoichiometry plays a critical role on the development of ferroelectricity.[29] Thus, the fact that DWs are gradually disappearing upon increasing the off-stoichiometry could also be due to the PbTiO$_3$ film losing its ferroelectric character with an increase in the concentration of point defects as the film composition deviates more and more from stoichiometric.[29] In order to elucidate the polar character of this same film, Raman spectroscopy was performed (see the experimental details in the Suppl. Info.). Raman spectra in a similar compound (a La$_x$Pb$_{1-x}$TiO$_3$ solid solution) shows that upon increasing the concentration of La, the E(TO) transverse optical phonon modes shift to lower wavenumber values and gradually vanish, such that for x $\geq$ 0.25 these Raman peaks disappear completely.[31] This is consistent with the ferroelectric-to-paralectric phase transition of PbTiO$_3$: in the high-temperature cubic phase (paralectric phase) Raman modes should disappear, whereas in the low-temperature tetragonal phase (ferroelectric phase) the A$_1$ and E modes (arising from the splitting of T$_{1u}$ and T$_{2u}$ modes within the tetragonal symmetry) become Raman active.[32] Figure S4a (Suppl. Info.) shows the Raman spectra measured on our film as a function of the [Ti]/[Ti]$_{Stoich}$ ratio. As observed, both the E(2TO) and A$_1$(2LO) phonon modes are present, even in the most off-stoichiometric regions of the film. Moreover, the wavenumber values of these phonon modes remain invariant within the range of [Ti]/[Ti]$_{Stoich}$ ratios explored in this work (Figures S4b in the Suppl. Info). Therefore, the gradual disappearance of the DWs in the increasingly off-stoichiometric regions of the film is not due to the loss of their polar character, but because of a preferential orientation of the polarization of the *c*-domains driven by the different [Pb]/[Ti] ratios in the PbTiO$_3$ film.

Thus, the non-trivial variation of the thermal conductivity found in the PbTiO$_3$ film deposited in a composition-spread geometry (Figure 1, bottom panel) is due to the combination of the two types of phonon scattering mechanisms: DWs and point defects. As the stoichiometric film presents the highest density of DWs, a minimum in $\kappa$ is found. When departing from [Pb]/[Ti] ≈ 1, DWs gradually vanish causing the rapid increase of $\kappa$, in spite of the increasing concentration of point defects. This shows the dominant role of DWs over point defects on phonon scattering, at least at moderate departures from stoichiometry. Further increase of the concentration of defects makes $\kappa$ decrease again, despite the complete absence of domain walls. Thus, in the most off-stoichiometric regions of the film the phonon scattering mechanism is totally dominated by point defects. Our findings reveal a gradual transition from a regime where the phonon scattering is governed by the DWs towards another one in which point defects prevail.

In order to disentangle both effects, we measured the temperature dependence of the cross-plane thermal conductivity and fitted the experimental data to the Debye model.[23,33] This is a crude approximation, in which the full phonon spectrum of the lattice is parametrized by the Debye temperature, $\Theta_D$, without any distinction of the different normal modes and with a single phonon velocity, $v_m$. This results in the following equation for the temperature dependent thermal conductivity:[34,35]

$$\kappa(T) = \frac{1}{2 v_m \pi} \int_0^{\omega_D} \tau(\omega) \frac{\hbar \omega^4}{k_B T^2} \frac{e^{\frac{\hbar \omega}{k_B}}}{\left(e^{\frac{\hbar \omega}{k_B}} - 1\right)^2} d\omega \qquad (1)$$

where $\omega_D$ the Debye's frequency, and $\tau(\omega)$ is the relaxation time. In this model the different phonon scattering processes are included additively into the relaxation time (Matthiessen's rule):

$$\tau^{-1}(\omega) = A + B\omega^2 T e^{-\frac{\Theta_D}{\alpha T}} + C\omega^4 = \frac{v_m}{\phi} + \frac{\hbar\gamma^2}{\bar{M}v^2\Theta_D} + \frac{3\delta^3}{\pi v_m^3}\sum_{i=1}^{N} x_i S_i^2 \omega^4 \qquad (2)$$

The first term accounts for the effect of boundary scattering, which, in the cross-plane thermal conductivity of our film, should be dominated by the total film thickness and the presence of ferroelectric DWs. The parameter $\phi$ (Casimir length), is the length a phonon can travel before it finds a boundary impeding it.[36] The second term accounts for the Umklapp process ($\gamma$ is the Grüneisen parameter, $\bar{M}$ is the average molar mass per atom, and $\alpha$ is a fitting parameter, which we kept $\alpha \approx 2$ as this parameter tends to lie between 1 and 3 in a large number of single-crystal materials[34,37–39]). The third term refers to the point defect (Rayleigh) scattering. The intentionally introduced cationic defects in the film deposited in a composition-spread geometry for this study are the main contributors to this term. Here $\delta^3$ is the average volume per atom, $x_i$ is the concentration of a particular type of defect, and $S_i$ is a parameter that encompasses all other effects that may affect the phonon propagation, like the change in mass, binding force, or the strain field around the vacancy itself.[23,33,40,41]

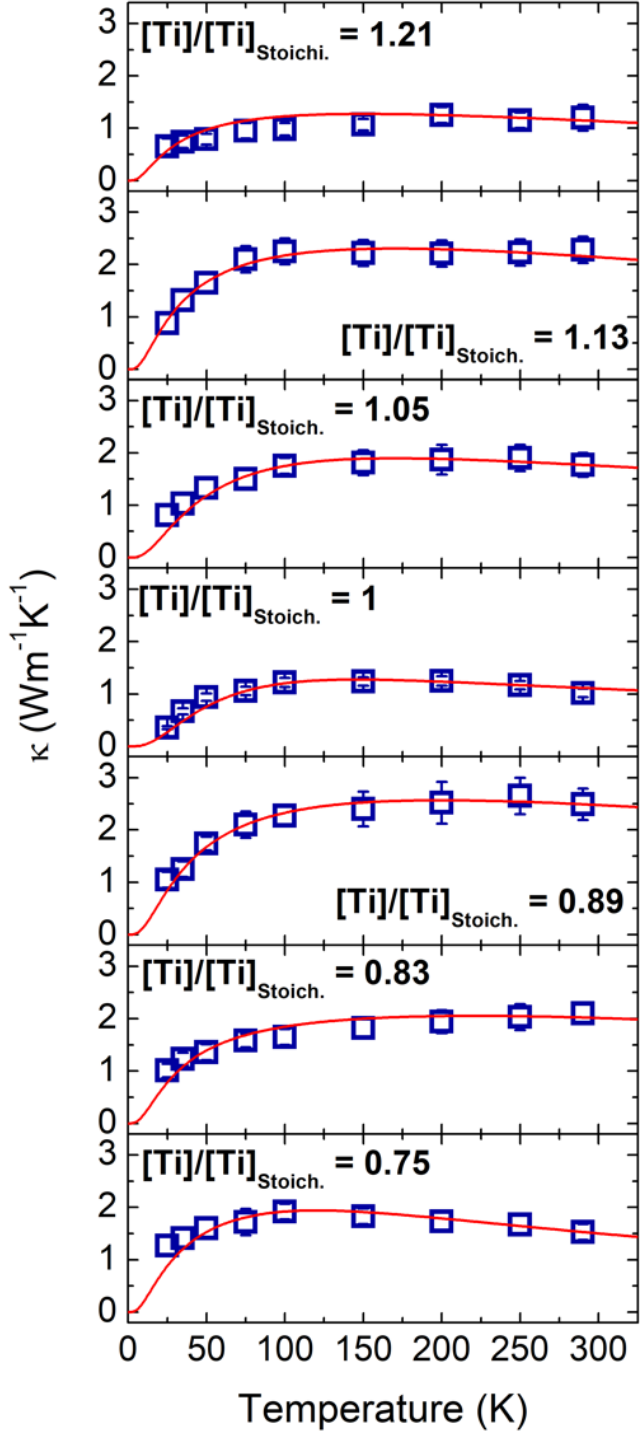

**Figure 3.** Temperature dependence of the thermal conductivity of a PbTiO$_3$ film deposited in a composition-spread geometry. The [Ti]/[Ti]$_{stoich.}$ ratio of the film is indicated in each panel, i.e., larger/smaller than one for titanium rich/poor pieces. The red solid lines are the best fit to the experimental thermal conductivity values (open symbols) using the Debye model (see text and equations (1) and (2)).

The experimental results and the fittings to equations (1)-(2) are shown in Figure 3. To minimize the number of free parameters in (2), we calculated their initial values as follows: for $x_i$ we used the values of vacancies obtained from EDS analysis. Other characteristics of the type of vacancy (size, mass, etc.) were also explicitly considered (see the Supp. Info.). The initial value of the Umklapp parameter was obtained from the stoichiometric piece ($B=2.4\times10^{-17}$ s K$^{-1}$) and used as a starting fitting value. The phonon mean velocity was also kept in the range $v_m = 3600 \pm 300$ m/s for the entire film (similar to the theoretical value $v_m \approx 3800$ m/s [Ref. 42]). For the boundary scattering term, $\phi$, we started our fittings with an initial value around 80 nm, the total film thickness. We then allowed $\phi$, $B$, and $C$ to vary slowly from their initial values until reaching the best agreement with the experimental data was achieved. We tested the robustness of our fittings against correlation between the different parameters by changing some of them randomly. The values of the fitting parameters are listed in the Suppl. Info.

Good fits to the experimental data can be obtained by changing the Umklapp term (without any trend) around its initial value (see the actual values from the fittings in Table S1 in the Suppl. Info.) and using $C$ that deviates only slightly from the initial calculation according to the experimental values of the cationic [Pb]/[Ti] ratio. In contrast, the boundary scattering term $\phi$ shows a clear dependence upon the composition of the film, with the smallest $\phi$ occurring for the stoichiometric region (Figure 4, top panel). The boundary size increases very rapidly with departure from stoichiometry, until reaching the limiting value of the film thickness. This is consistent with the gradual disappearance of the domain walls according to the PFM and X-ray analyses discussed above.

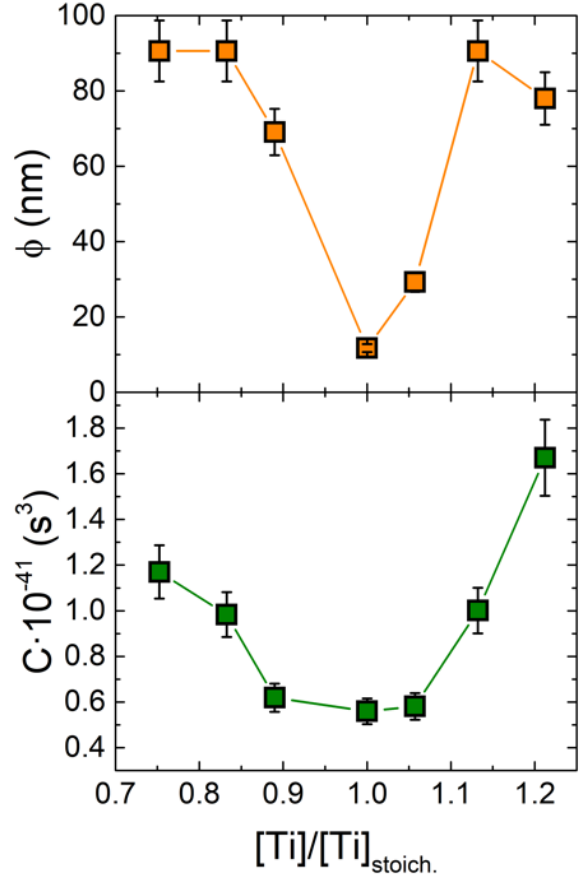

**Figure 4.** Boundary size, $\phi$ (top panel), and point-defect-scattering term, $C$ (bottom panel), as a function of the titanium concentration ($[Ti]/[Ti]_{stoich.}$) in the PbTiO$_3$ film deposited in a composition-spread geometry.

It is worth noting that the defect-scattering parameter, $C$, does not vary significantly for small departures from stoichiometry. This explains why, at low/moderate concentrations of point defects, DWs are the main phonon scatterers, playing the major role in determining the $\kappa$ of ferroelectric materials. Yet, on further increasing the off-stoichiometry, the value of $C$ grows rapidly, while $\phi$ remains approximately constant, limited by the film thickness (Figure 4). Thus, at large concentrations of cationic vacancies, point-defect scattering determines the thermal conductivity of the PbTiO$_3$ film. The consequence of the gradual transition between the dominance of both scattering mechanisms is that $\kappa$ presents a maximum at a moderate concentration of point defects.

**CONCLUSIONS**

Our results demonstrate that DWs are the main phonon scatterers in ferroelectric PbTiO$_3$, not only for the stoichiometric scenario, but also when the presence of cation defects is significant ($\approx$15%). Thus, our findings reveal the effectiveness of reconfigurable DWs in reducing the thermal conductivity, clearly surpassing the effects of point defects when their concentration is moderate. The significant potential of ferroelectric materials to control phonon transport in solid-state devices is hence substantiated.

ASSOCIATED CONTENT

**Supporting Information**. Additional information is provided on the epitaxial growth and structural characterization of the PbTiO$_3$ thin film deposited in a composition-spread geometry, piezoresponse force microscopy experiments, Raman spectra, thermal conductivity measurements, and Debye model.

AUTHOR INFORMATION

**Corresponding Author**

*E-mail: eric.langenberg@ub.edu and f.rivadulla@usc.es

**Notes**

The authors declare no competing financial interest.

ACKNOWLEDGMENT

This work has received financial support from Ministerio de Economía y Competitividad (Spain) under projects No. MAT2016-80762-R, PID2019-104150RB-100 and PGC2018-097789-B-I00, Xunta de Galicia (Centro singular de investigación de Galicia accreditation 2016-2019, ED431G/09), the European Union (European Regional Development Fund-ERDF) and the European Commission through the Horizon H2020 funding by H2020-MSCA-RISE-2016-Project No. 734187–SPICOLOST. E.L. is a Serra Húnter Fellow (Generalitat de Catalunya). E.L. acknowledges the funding received from the European Union's Horizon 2020 research and innovation program through the Marie Skłodowska-Curie Actions: Individual Fellowship-Global Fellowship (Ref. MSCA-IF-GF-708129). D.B. acknowledges financial support from MINECO (Spain) through an FPI fellowship (BES-2017-079688). The work at Cornell was supported by the Army Research Office under grant W911NF-16-1-0315. H.P. acknowledges support from the National Science Foundation [Platform for the Accelerated Realization, Analysis, and Discovery of Interface Materials (PARADIM)] under Cooperative Agreement No. DMR-1539918.

Supporting Information

# Deconvolution of phonon scattering by ferroelectric domain walls and point defects in a PbTiO$_3$ thin film deposited in a composition-spread geometry

David Bugallo[1], Eric Langenberg[1,2,3,*], Elias Ferreiro-Vila[1], Eva H. Smith[2], Christina Stefani[4], Xavier Batlle[3], Gustau Catalan[4], Neus Domingo[4], Darrell G. Schlom[2,5,6], and Francisco Rivadulla[1,*]

[1]CiQUS-Universidade de Santiago de Compostela, Santiago de Compostela 15782, Spain

[2]Department of Materials Science and Engineering, Cornell University, Ithaca, New York 14853, USA

[3]Department de Física de la Matèria Condensada and Institut de Nanociència i Nanotecnologia (IN2UB), Universitat de Barcelona, Barcelona 08028, Spain.

[4]Catalan Institute of Nanoscience and Nanotechnology (ICN2), CSIC, Barcelona Institute of Science and Technology, Campus Universitat Autònoma de Barcelona, Bellaterra, 08193 Barcelona, Spain

[5]Kavli Institute at Cornell for Nanoscale Science, Ithaca, New York 14853, USA

[6]Leibniz-Institut für Kristallzüchtung, Max-Born-Str. 2, 12489 Berlin, Germany

1. EXPERIMENTAL SECTION

**Growth of the PbTiO$_3$ film**. The PbTiO$_3$ thin film was deposited in a composition-spread geometry by reactive molecular-beam epitaxy (MBE) in a Veeco GEN 10 system using distilled ozone as an oxidant and elemental lead and titanium as source materials. Lead is supplied from a conventional MBE effusion cell, whereas titanium is sublimed from a Ti-Ball™ (Varian Associates, Vacuum Products Division).[1] During growth lead and titanium were continuously codeposited, achieving phase-pure PbTiO$_3$ in the growing film by adsorption-control.[2] Typical lead and titanium fluxes were $1.7 \times 10^{14}$ and $1.5 \times 10^{13}$ atoms/(cm$^2 \cdot$s), respectively. Distilled ozone, which consists of about 80% O$_3$ and 20% O$_2$, was used as the oxidant.[3] The background pressure of distilled ozone during growth and the substrate temperature were fixed at $9 \times 10^{-6}$ Torr and 600 °C, respectively. The gradient in composition across the film surface was achieved by depositing the film onto a 1"-diameter (001)-oriented SrTiO$_3$ single-crystal substrate that purposefully was not rotated during growth. In this way a continuous variation of the [Pb]/[Ti] ratio along one of the diameters of the wafer is obtained due to the fact that the lead and the titanium sources are placed nearly diametrically opposite one to each other (see Figure S1). After growth the sample was cut into seven pieces along this diameter. Each of the resulting PbTiO$_3$ pieces has a different [Pb]/[Ti] ratios: three increasingly lead rich, three increasingly titanium rich, and one stoichiometric.

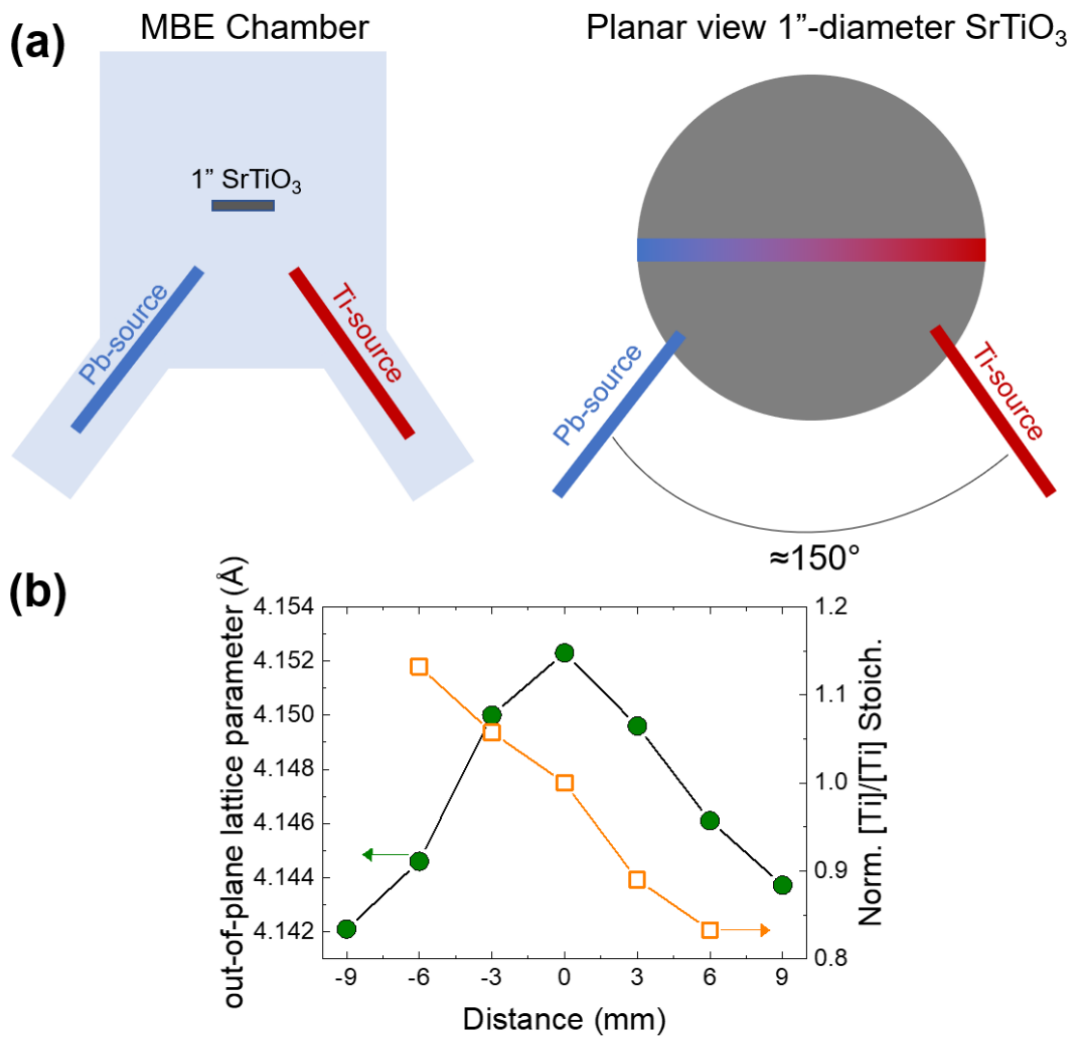

**Figure S1.** (a) Scheme of the placement of the substrate, and titanium and lead sources inside the MBE chamber. (b) Out-of-plane lattice parameter and [Ti]/[Ti]$_{Stoich.}$ at different positions

along the diameter where the in-plane gradient in composition take place. Distance 0 corresponds to the center of the wafer; ± 9 mm correspond to almost the extremes of the substrate.

**Compositional characterization**. The [Pb]/[Ti] composition ratio was determined by energy dispersive spectroscopy (EDS) analysis using an environmental scanning Electron microscope (EVO LS15, ZEISS) with an INCA-X act detector (Oxford). We worked at low voltages (7 kV) to diminish the electron beam penetration into the substrate and thus maximize the interaction with the film. We express the difference in concentration of the titanium cation in the different samples, after having normalized it by the concentration of strontium, which is only due to the substrate. This change is directly related to the ratio between lead and titanium in the samples (a higher amount of titanium means a lower amount of lead), which was confirmed by structural characterization performed by X-ray diffraction (XRD).

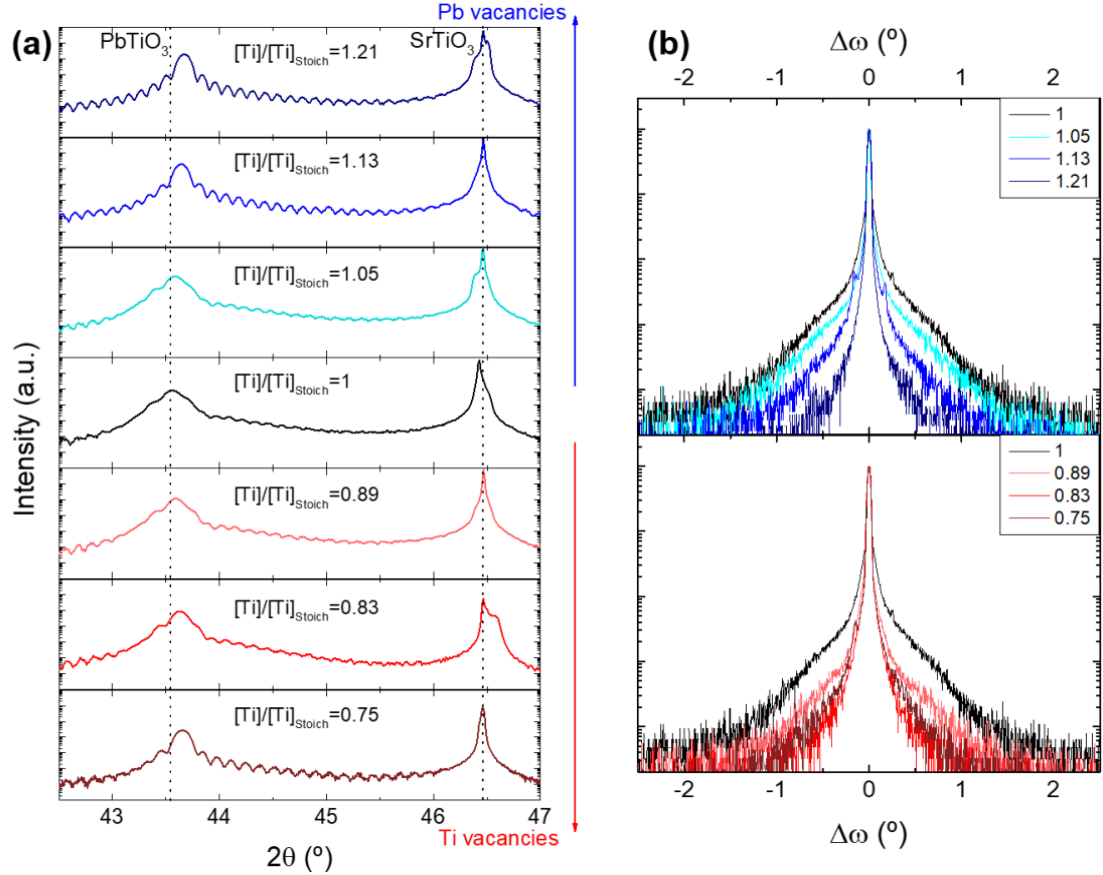

**Figure S2. (a)** XRD θ-2θ scans around the 002 reflection of both the SrTiO$_3$ substrate and the PbTiO$_3$ pieces. The vertical dotted lines indicate the position of the 002 reflection peak (SrTiO$_3$ substrate and PbTiO$_3$ film) of the stoichiometric piece. **(b)** Rocking curves (Δω) around the 2θ value of the 001 reflection of the PbTiO$_3$ film. The value of the ratio [Ti]/[Ti]$_{Stoich.}$ is indicated at the top center and at the top right corner of the measurements, for panels (a) and (b), respectively. Note that the intensity is plotted on a logarithmic scale.

**Structural characterization.** The structure of the films was characterized by XRD measurements using a four-circle diffractometer (PANalytical Empyrean) with a Ge 220 double bounce monochromator and a PIXcel$^{3D}$ area detector, measuring θ-2θ scans, rocking curves (Figure S2a and S2b, respectively) and reciprocal space maps (RSM) of each piece of the PbTiO$_3$ film (Figure S3). The diffraction peak of the PbTiO$_3$ film was observed to become narrower in the RSM the more off composition the films was (Figure S3). The same trend was observed in the rocking curves (Figure S2b).

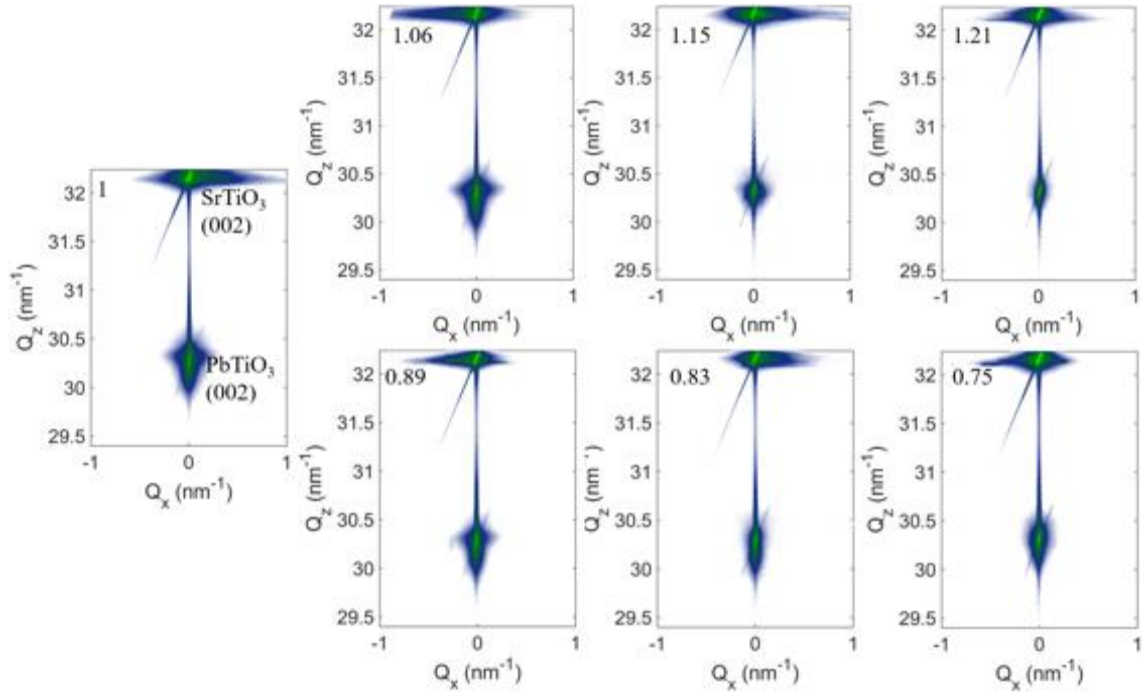

**Figure S3.** Reciprocal space maps around the 002 reflection of both the SrTiO$_3$ substrate and the PbTiO$_3$ film. The value for the ratio [Ti]/[Ti]$_{Stoich.}$ is indicated in the top left corner of each image.

**Ferroelectric domain characterization**. The ferroelectric domain patterns were assessed by piezoresponse force microscopy (PFM) in dual *ac* resonance track (DART) mode, using an Asylum MFP-3D Bio (Asylum Research). The typical excitation voltage was in the range of $V_{ac}$ = 1.5 - 3.0 volts. PtIr conducting tips (PointProbe® Plus-Electrostatic Force Microscopy from Nanosensors) were employed, with typical resonance frequencies around 350 kHz for the vertical piezoresponse.

**Polar state**. In order to corroborate if the films with the highest concentration of defects are still polar, Raman spectroscopy was performed using a HeCd laser with a wavelength of 325 nm. The system is equipped with a dispersive spectrometer Jobin-Yvon LabRam HR 800, coupled to an optical microscope Olympus BXFM. The CCD detector was cooled at -70 ºC, and the dispersive grating has 2400 lines/mm. The laser power used in the measurements was 0.5 mW.

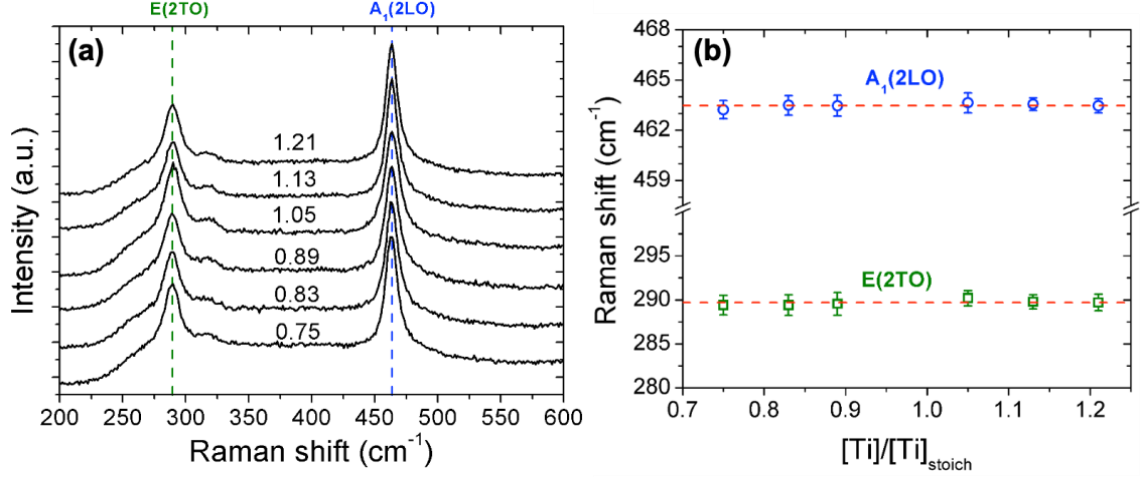

**Figure S4. (a)** Raman spectra of the pieces of the $PbTiO_3$ film with different $[Ti]/[Ti]_{Stoich}$ ratios (indicated on top of each Raman spectrum). **(b)** Wavenumber of the $A_1(2LO)$ and $E(2TO)$ phonon modes as a function of the $[Ti]/[Ti]_{Stoich}$ ratios. The dashed red line is the average wavenumber value of the each phonon mode.

The Raman spectra of the $PbTiO_3$ pieces displaying different $[Ti]/[Ti]_{stoich.}$ ratios are shown in Figure S4a. The evolution of the wavenumber of the $E(2TO)$ and $A_1(2LO)$ phonon modes as a function of the $[Ti]/[Ti]_{stoich.}$ ratio is plotted in Figure S4b. The center of these Raman peaks was computed using Gaussian functions.

**Thermal Conductivity**. The temperature dependence of the cross-plane thermal conductivity of each sample was measured by the 3ω method[4,5] between 20 K and 300 K. The details of this method are explained below.

A narrow line of Au (100 nm thick, 10 μm width, 1 mm length with 10 nm of Cr for adhesion) is deposited on top of the sample using evaporation and optical lithography. This metal resistor serves as a heater and a thermometer simultaneously. By applying an *ac* current at frequency ω and amplitude $I_0$ through the resistor, a temperature oscillation at $2\omega$ ($\Delta T_{2\omega}$) is produced due to conventional Joule heating. The amplitude of this oscillation is indirectly measured through the third harmonic voltage, $V_{3\omega}$, from the formula given below (a detailed description can be found in Ref. 6,7):

$$V_{3\omega} = \frac{I_0}{2}\left(\frac{dR}{dT}\right)\Delta T_{2\omega}.$$

The heating produced at frequency $2\omega$, which depends on the frequency, $\omega$, of the applied *ac* current, is related to the thermal properties of the medium underneath the resistor. Specifically, by solving the one-dimensional heat equation, the following expression is obtained:[8]

$$\Delta T_{2\omega} = \frac{P}{\pi l \kappa}\left[\frac{1}{2}\ln\left(\frac{D}{\left(\frac{w}{2}\right)^2}\right) - \frac{1}{2}\ln(2\omega) + \eta - \frac{i\pi}{4}\right],$$

where $P$ ($=I_0^2 R$ with $R$ the electrical resistance of the metal line) is the power dissipated through the resistor, $l$ is its length, $\kappa$ and $D$ are, respectively, the thermal conductivity and the thermal diffusivity of the material immediately beneath the resistor, $w$ is the length of the resistor, and $\eta$ is a parameter that depends on the material used,[5,9,10] in particular, on the thermal diffusivity as we have previously shown.[6] Note that in order to consider the heat flow as being one-dimensional, the width of the resistor, $w$, should be much smaller than the penetration length of the thermal wave.[8]

When the sample underneath the resistor consists of a thin film and a substrate, the solution of the heat equation is slightly different. As long as the thickness of the film is much smaller than the width of the resistor ($t \ll w$), which is our case, the transmission of heat through the film can be considered one-dimensional. In this scenario, the following expression is deduced (Ref. 5):

$$\Delta T_{2\omega} = \frac{P}{\pi l \kappa}\left[\frac{1}{2}\ln\left(\frac{D}{\left(\frac{w}{2}\right)^2}\right) - \frac{1}{2}\ln(2\omega) + \eta - \frac{i\pi}{4}\right] + \frac{P}{\kappa_{film}}\frac{t}{wl} = \Delta T_{substrate} + \Delta T_{film},$$

where the second term in the equation is the temperature increase produced by the film. At each frequency, this contribution of the film appears as an offset in comparison with the heating produced when just the substrate is present (Figure S5a). Then by subtraction of the heating produced in the substrate ($\Delta T_{substrate}$) to that obtained when measuring the film + substrate ($\Delta T_{film+substrate}$), the thermal conductivity of the thin film can be obtained from:

$$\kappa_{film} = \frac{P}{\Delta T_{film}}\frac{t}{wl}.$$

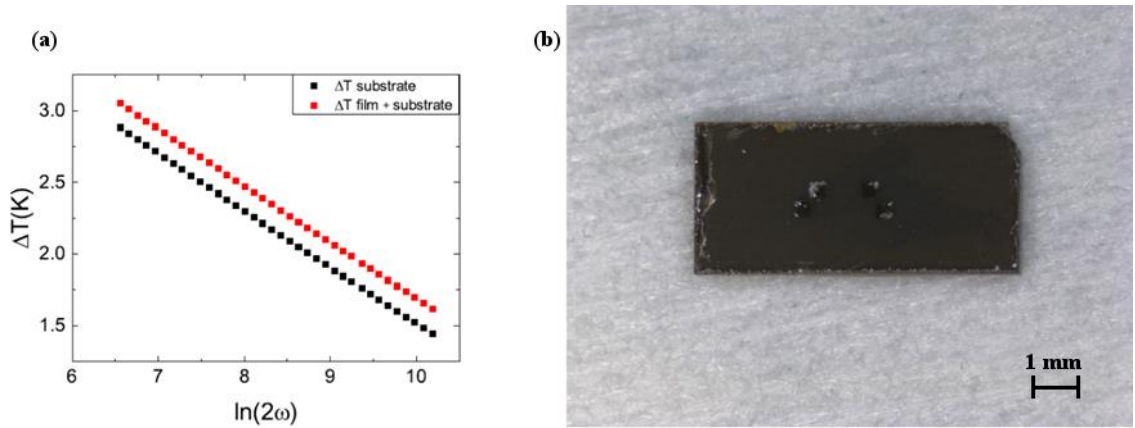

**Figure S5.** (a) $\Delta T$ of the stoichiometric PbTiO$_3$ piece and the SrTiO$_3$ substrate measured at 250 K by dissipating 25 mW through the resistor. (b) Optical photograph of one of the pieces of the PbTiO$_3$ film with the resistor deposited on top of it.

The measurements were performed by applying current at frequencies from 56 to 2120 Hz to produce a heat dissipation of 15 and 25 mW at each temperature (from 25 to 290 K) for all pieces of the PbTiO$_3$ film and also to a bare SrTiO$_3$ substrate with a Keithley 6221 AC source. The voltage drop at $3\omega$ was measured with a Stanford Research Systems SR830 lock-in amplifier. As the ratio $V_{1\omega}/V_{3\omega}$ is around $10^3$, a circuit was used to cancel out the voltage at $1\omega$, which is described in Ref. 6.

## 2. DEBYE MODEL

In the definition of the relaxation time, the term $S_i^2$ has three contributions that arise, respectively, from differences in mass of the defect, $S_1$; the change in the bonds around it, $S_2$; and the change in the radius, $S_3$. The form that they have is the following:

$$S_i^2 = S_1^2 + (S_2 + S_3)^2 = \frac{1}{12}\left(\frac{\Delta M}{\overline{M}}\right)^2 + \frac{2}{3}\left(1 + \frac{2}{\sqrt{3}}Q + \frac{1}{3}Q^2\right)\gamma^2\left(\frac{\Delta R}{\overline{R}}\right)^2,$$

where $\Delta M$ ($\Delta R$) is the change of mass (radius) caused by the defect with respect to the average, and $Q$ is a factor depending on the harmonicity of the bonds according to Klemens.[11]

Note that this approach is based on the assumption that every position in the crystal is the same. In reality, since there is a big difference in both mass and radius between the distinct ions forming the material, the contributions for the different atoms can be separated, allowing a better description of the system to be achieved. Consequently, in a system $A_aB_bC_c$ we can divide the lattice in three sublattices, for lead, titanium, and oxygen atoms, respectively.[12,13]

$$S_i^2 = S_1^2 + (S_2 + S_3)^2 = \frac{a}{a+b+c}\left[\frac{1}{12}\left(\frac{M_A}{\overline{M}}\right)^2 + \frac{2}{3}\left(1 + \frac{2}{\sqrt{3}}Q + \frac{1}{3}Q^2\right)\gamma^2\left(\frac{R_A}{\overline{R}}\right)^2\right] +$$

$$\frac{b}{a+b+c}\left[\frac{1}{12}\left(\frac{M_B}{\overline{M}}\right)^2 + \frac{2}{3}\left(1 + \frac{2}{\sqrt{3}}Q + \frac{1}{3}Q^2\right)\gamma^2\left(\frac{R_B}{\overline{R}}\right)^2\right] + \frac{c}{a+b+c}\left[\frac{1}{12}\left(\frac{M_C}{\overline{M}}\right)^2 + \frac{2}{3}\left(1 + \frac{2}{\sqrt{3}}Q + \frac{1}{3}Q^2\right)\gamma^2\left(\frac{R_C}{\overline{R}}\right)^2\right].$$

In the presence of vacancies, the value of the defect's mass is zero. In addition, when there is a deficit of titanium, since lead can also be present with +4 valence, the possibility of lead occupying the titanium position was taken into account, using its mass and radius to calculate the point defect scattering parameter for the titanium-deficient samples. In that case as exposed by Klemens, the value of $Q$ used is 4.2, as the harmonicity of the nearest linkages is kept; while in the case of having vacancies the value is 3.2 as the harmonicity would change.[11] Furthermore, a 5% concentration of oxygen vacancies was included in the calculation for all of the samples.

In table S1 the parameters from the fit of all of the samples are listed, aside from the $\alpha$ parameter in the exponential for the Umklapp scattering, which was kept at 2 for all of the samples. It is important to note that the parameter $A$ was limited to not be greater than $\approx 5 \times 10^{10}$ s$^{-1}$ as that value corresponds to approx. 100% of the film thickness.

**Table S1.** Parameters obtained from the Debye Model fit of the different samples: Boundary scattering $A$, Umklapp scattering $B$, and Point Defect scattering $C$.

| Sample [Ti]/[Ti]$_{Stoich}$ | $A$(s$^{-1}$) | $B$(s·K$^{-1}$) | $C$(s$^3$) |
|---|---|---|---|
| 0.75 | 4.3·10$^{10}$ | 8.0·10$^{-18}$ | 1.2·10$^{-41}$ |
| 0.83 | 5.3·10$^{10}$ | 3.0·10$^{-18}$ | 9.8·10$^{-42}$ |
| 0.89 | 6.5·10$^{11}$ | 4.0·10$^{-18}$ | 6.19·10$^{-42}$ |
| 1 | 3.3·10$^{11}$ | 2.4·10$^{-17}$ | 5.6·10$^{-42}$ |
| 1.05 | 1.3·10$^{11}$ | 7.8·10$^{-18}$ | 5.8·10$^{-42}$ |
| 1.13 | 4.3·10$^{10}$ | 4.0·10$^{-18}$ | 1.0·10$^{-41}$ |
| 1.21 | 5.0·10$^{10}$ | 5.3·10$^{-18}$ | 1.7·10$^{-41}$ |